\documentclass{mymod}

\newcommand{ \be}{\begin{equation}}
\newcommand{ \ee}{\end{equation}}
\newcommand{ \bea}{\begin{eqnarray}}
\newcommand{ \eea}{\end{eqnarray}}
\newcommand{ \mysmall}[1]{\scriptscriptstyle #1} 

\newcommand{ \at}{a_{\tau}}
\newcommand{ \mw}{M_{\mysmall{W}}}
\newcommand{ \mz}{M_{\mysmall{Z}}}
\newcommand{ \mh}{M_{\mysmall{H}}}

\newcommand{ \eq}[1]{Eq.~(\ref{eq:#1})}
\newcommand{ \gev}  {\mbox{ GeV}}

\begin{document}

\markboth{S.\ Eidelman \& M.\ Passera}{Theory of the $\tau$ lepton anomalous
magnetic moment}

\title{THEORY OF THE {\large $\tau$} LEPTON ANOMALOUS MAGNETIC MOMENT}

\author{S.\ EIDELMAN}
\address{ Budker Institute of Nuclear Physics\\
  11 Academician Lavrentiev prospect,
  Novosibirsk, 630090, Russia\\
  eidelman@inp.nsk.su}

\author{M.\ PASSERA}

\address{Dipartimento di Fisica ``G.~Galilei'', Universit\`{a}
        di Padova and \\ INFN, Sezione di Padova, Via Marzolo 8,
        I-35131, Padova, Italy \\
        passera@pd.infn.it}

\maketitle

\begin{abstract}
%
This article reviews and updates the Standard Model prediction of the $\tau$
lepton $g$$-$$2$. Updated QED and electroweak contributions are presented,
together with new values of the leading-order hadronic term, based on the
recent low energy $e^+ e^-$ data from BaBar, CMD-2, KLOE and SND, and of the
hadronic light-by-light contribution. The total prediction is confronted to
the available experimental bounds on the $\tau$ lepton anomaly, and
prospects for its future measurements are briefly discussed.

\end{abstract}

\section{Introduction}	
\label{sec:INTRO}

Numerous precision tests of the Standard Model ({\small SM}) and searches
for its possible violation have been performed in the last few decades,
serving as an invaluable tool to test the theory at the quantum level. They
have also provided stringent constraints on many ``New Physics'' ({\small
NP}) scenarios.
A typical example is given by the measurements of the anomalous magnetic
moment of the electron and the muon, where recent experiments reached the
fabulous relative precision of 0.7 ppb~\cite{odom} and 0.5 ppm,\cite{bnl}
respectively. These experiments measure the so-called gyromagnetic factor
$g$, defined by the relation between the particle's spin $\vec{s}$ and its
magnetic moment $\vec{\mu}$,
\be
\vec{\mu}=g \frac {e} {2m} \vec{s},
\ee
where $e$ and $m$ are the charge and mass of the particle.  In the Dirac
theory of a charged point-like spin-$1/2$ particle, $g=2$.  Quantum
Electrodynamics ({\small QED}) predicts deviations from Dirac's value, as
the charged particle can emit and reabsorb virtual photons.  These {\small
QED} effects slightly increase the $g$ value. It is conventional to express
the difference of $g$ from 2 in terms of the value of the so-called anomalous
magnetic moment, a dimensionless quantity defined as $a = (g - 2)/2$.

The anomalous magnetic moment of the electron, $a_e$, is rather insensitive
to strong and weak interactions, hence providing a stringent test of {\small
QED} and leading to the most precise determination of the fine-structure
constant $\alpha$ to date.\cite{Gabrielse_a_2006,MP06} On the other hand,
the $g$$-$$2$ of the muon, $a_{\mu}$, allows to test the entire {\small SM},
as each of its sectors contributes in a significant way to the total
prediction. Compared with $a_e$, $a_{\mu}$ is also much better suited to
unveil or constrain {\small NP} effects. Indeed, for a lepton $l$, their
contribution to $a_l$ is generally expected to be proportional to
$m_l^2/\Lambda^2$, where $m_l$ is the mass of the lepton and $\Lambda$ is
the scale of {\small NP}, thus leading to an $(m_{\mu}/m_e)^2 \sim 4\times
10^4$ relative enhancement of the sensitivity of the muon versus the
electron anomalous magnetic moment. This more than compensates the much
higher accuracy with which the $g$ factor of the latter is known.  The
anomalous magnetic moment of the $\tau$ lepton, $a_{\tau}$, would suit even
better; however, its direct experimental measurement is prevented by the
relatively short lifetime of this lepton, at least at present. The existing
limits are based on the precise measurements of the total and differential
cross sections of the reactions $e^+e^- \to e^+e^-\tau^+\tau^-$ and $e^+e^-
\to Z \to \tau^+\tau^-\gamma$ at {\small LEP} energies. The most stringent
limit, $-0.052 < a_{\tau} < 0.013$ at 95\% confidence level, was set by the
{\small DELPHI} collaboration,\cite{delphi} and is still more than an order
of magnitude worse than that required to determine $a_{\tau}$.

In the 1990s it became clear that the accuracy of the theoretical prediction
of the muon $g$$-$$2$, challenged by the E821 experiment underway at
Brookhaven,\cite{bnl} was going to be restricted by our knowledge of its
hadronic contribution.  This problem has been solved by the impressive
experiments at low-energy $e^+e^-$ colliders, where the total hadronic cross
section (as well as exclusive ones) were measured with high precision,
allowing a significant improvement of the uncertainty of the leading-order
hadronic contribution.\cite{se06,kaoru06} As a result, the accuracy of the
{\small SM} prediction for $a_{\mu}$ now matches that of its measurement.
In parallel to these efforts, very many improvements of all other 
sectors of the
{\small SM} prediction were carried on by a large number of theorists (see
Refs.~\refcite{g-2_mureviews,MP04} for reviews). All these experimental and
theoretical developments allow to significantly improve the theoretical
prediction for the anomalous magnetic moment of $\tau$ lepton as well.

In this article we review and update the {\small SM} prediction of
$a_{\tau}$, analyzing in detail the three contributions into which it is
usually split: {\small QED}, electroweak ({\small EW}) and hadronic. Updated
{\small QED} and {\small EW} contributions are presented in
Secs.~\ref{sec:QED} and \ref{sec:EW}; new values of the leading-order
hadronic term, based on the recent low energy $e^+ e^-$ data from BaBar,
{\small CMD-2}, {\small KLOE} and {\small SND}, and of the hadronic
light-by-light contribution are presented in Sec.~\ref{sec:HAD}. The total
{\small SM} prediction is confronted to the available experimental bounds on
the $\tau$ lepton $g$$-$$2$ in Sec.~\ref{sec:SM}, and prospects for its
future measurements are briefly discussed in Sec.~\ref{sec:CONC}, where
conclusions are drawn.

\section{QED Contribution to $\at$}
\label{sec:QED}

The {\small QED} part of the anomalous magnetic moment of the $\tau$ lepton
arises from the subset of {\small SM} diagrams containing only leptons and
photons. This dimensionless quantity can be cast in the general
form:\cite{KM90}
\be
    a_{\tau}^{\mysmall \rm QED} = A_1 +
                   A_2 \left( \frac{m_{\tau}}{m_e} \right) +
                   A_2 \left( \frac{m_{\tau}}{m_{\mu}} \right) +
                   A_3 \left( \frac{m_{\tau}}{m_e},
		                \frac{m_{\tau}}{m_{\mu}}\right),
\label{eq:atauqedgeneral}
\ee
where $m_e$, $m_{\mu}$ and $m_{\tau}$ are the electron, muon and $\tau$
lepton masses, respectively.  The term $A_1$, arising from diagrams
containing only photons and $\tau$ leptons, is mass and flavor independent.
In contrast, the terms $A_2$ and $A_3$ are functions of the indicated mass
ratios, and are generated by graphs containing also electrons and/or muons.
The functions $A_i$ ($i=1,2,3$) can be expanded as power series in
$\alpha/\pi$ and computed order-by-order
\be
    A_i = A_i^{(2)}\left(\frac{\alpha}{\pi} \right)
    + A_i^{(4)}\left(\frac{\alpha}{\pi} \right)^{2}
    + A_i^{(6)}\left(\frac{\alpha}{\pi} \right)^{3}
    + A_i^{(8)}\left(\frac{\alpha}{\pi} \right)^{4} +\cdots.
\ee
Only one diagram is involved in the evaluation of the lowest-order
(first-order in $\alpha$, second-order in the electric charge) contribution
-- it provides the famous result by Schwinger $A_1^{(2)} =
1/2$.\cite{Sch48} The mass-dependent coefficients $A_2$ and $A_3$,
discussed below, are of higher order. They were derived using the latest
{\small CODATA}\cite{CODATA02} recommended mass ratios:
\bea
      m_{\tau}/m_e     &=& 3477.48 (57)
\label{eq:rte}\\
      m_{\tau}/m_{\mu} &=& 16.8183 (27).
\label{eq:rtm}
\eea
The value for $m_{\tau}$ adopted by {\small CODATA} in
Ref.~\refcite{CODATA02}, $m_{\tau}= 1776.99\, (29)$ MeV, is based on the
{\small PDG} 2002 result.\cite{PDG02} It remained unchanged until very
recently (see Refs.~\refcite{PDG04,PDG06}), when preliminary results of two
new measurements (from the Belle\cite{belle_mtau} and {\small
KEDR}\cite{kedr_mtau} detectors) were reported. The central values of the
new mass values are slightly lower than the current world average value, but
agree with it within the uncertainties, which are approaching that of the
world average value (used in this work).

\subsection{Two-loop Corrections}
\label{subsec:QED2}

\begin{figure}[th]
\centerline{\psfig{file=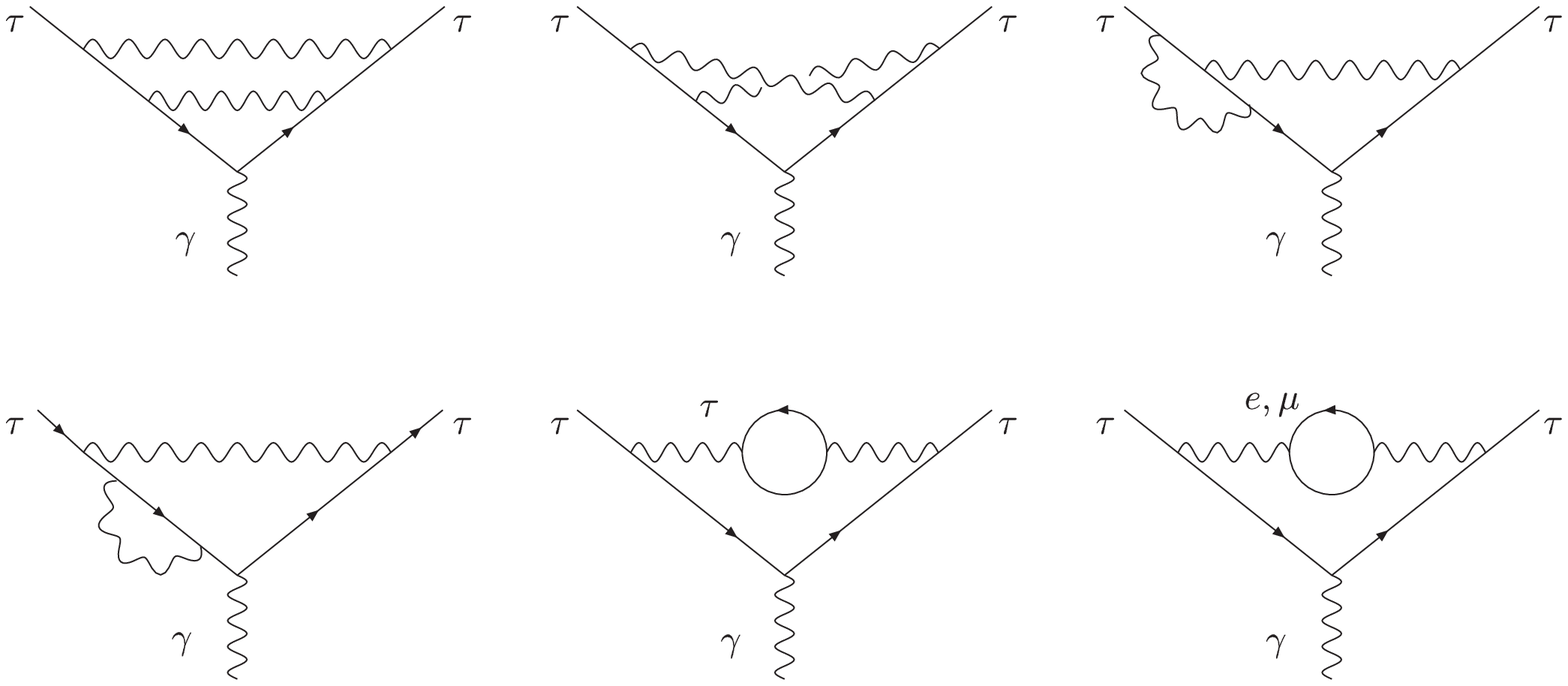,width=12.5cm}}
\vspace*{8pt}
\caption{The {\small QED} diagrams contributing to the $\tau$ lepton
         $g$$-$$2$ at order $\alpha^2$. The mirror reflections (not shown)
         of the third and fourth diagrams must be included as well.}
\label{fig:qed2}
\end{figure}

Seven diagrams contribute to the fourth-order coefficient $A_1^{(4)}$, one
to $A_2^{(4)}(m_{\tau}/m_e)$ and one to $A_2^{(4)}(m_{\tau}/m_{\mu})$. They
are depicted in Fig.~\ref{fig:qed2}. As there are no two-loop diagrams
contributing to $a_{\tau}^{\mysmall \rm QED}$ that contain both virtual
electrons and muons, $A_3^{(4)}(m_{\tau}/m_e,m_{\tau}/m_{\mu}) = 0$. The
mass-independent coefficient has been known for almost fifty
years:\cite{So57-58-Pe57-58}
\bea
    A_1^{(4)} &=& \frac{197}{144} + \frac{\pi^2}{12}
                + \frac{3}{4}\zeta(3) - \frac{\pi^2}{2} \ln2 \nonumber \\
                &=& -0.328 \, 478 \, 965 \, 579 \, 193 \, 78 \ldots,
\label{eq:A14}
\eea
where $\zeta(s)$ is the Riemann zeta function of argument $s$. 
For $l\!=\!e$, $\mu$ or $\tau$, the coefficient of the two-loop
mass-dependent contribution to $a_l^{\mysmall \rm QED}$, $A_2^{(4)}(1/x)$,
with $x=m_j/m_l$, is generated by the diagram with a vacuum polarization
subgraph containing the virtual lepton $j$.  This coefficient was first
computed in the late 1950s for the muon $g$$-$$2$ with $x = m_e/m_{\mu} \ll
1$, neglecting terms of $O(x)$.\cite{SWP57} The exact expression for $0<x<1$
was reported by Elend in 1966.\cite{El66} However, its numerical evaluation
was considered tricky because of large cancellations and difficulties in the
estimate of the accuracy of the results, so that common practice was to use
series expansions instead.\cite{Samuel91,Samuel93,CS99} Taking advantage of
the properties of the dilogarithm ${\rm Li}_2(z)=-\int_0^z dt
\ln(1-t)/t$,\cite{Lewin} the exact result was cast in Ref.~\refcite{MP04} in
a very simple and compact analytic form, valid, contrary to the one in
Ref.~\refcite{El66}, also for $x \geq 1$ (the case relevant to
$a_e^{\mysmall \rm QED}$ and part of $a_{\mu}^{\mysmall \rm QED}$):
\bea
     A_2^{(4)}(1/x)   &=&
     -\frac{25}{36} - \frac{\ln x}{3}
     +x^2 \left(4+3\ln x \right) +  \frac{x}{2} \left(1-5 x^2\right)
     \times
     \nonumber \\     &&
     \times\left[\frac{\pi^2}{2}
       - \ln x \, \ln \left( \frac{1-x}{1+x} \right)
       - {\rm Li}_2(x) + {\rm Li}_2(-x) \right] +
     \nonumber \\     &&
     +x^4 \left[ \frac{\pi^2}{3} -2\ln x \, \ln \left(\frac{1}{x}-x\right)
     -{\rm Li}_2(x^2)\right].
\label{eq:EA24}
\eea
For $x=1$, \eq{EA24} gives $A_2^{(4)}(1) = 119/36 - \pi^2/3$; of course,
this contribution is already part of $A_1^{(4)}$ in \eq{A14}. Numerical
evaluation of \eq{EA24} with the mass ratios given in
Eqs.~(\ref{eq:rte})--(\ref{eq:rtm}) yields the two-loop mass-dependent
{\small QED} contributions to the anomalous magnetic moment of the $\tau$
lepton,\cite{MP06}
\bea
     A_2^{(4)}(m_{\tau}/m_e)      & = & 2.024 \, 284 \, (55),
\label{eq:TA24e}
\\
     A_2^{(4)}(m_{\tau}/m_{\mu}) & = & 0.361 \, 652 \, (38).
\label{eq:TA24m}
\eea
These two values are very similar to those computed via a dispersive
integral in Ref.~\refcite{Narison01} (which, however, contain no estimates
of the uncertainties). Equations~(\ref{eq:TA24e}) and (\ref{eq:TA24m}) are
also in agreement (but more accurate) with those of Ref.~\refcite{Samuel93}.
Adding up Eqs.~(\ref{eq:A14}), (\ref{eq:TA24e}) and (\ref{eq:TA24m}) one
gets:\cite{MP06}
\be
       C_{\tau}^{(4)} = 2.057 \, 457 \, (93)
\label{eq:TC2}
\ee
(note that the uncertainties in $m_{\tau}/m_e$ and $m_{\tau}/m_{\mu}$ are
correlated). The resulting error $9.3 \times 10^{-5}$ leads to a $5 \times
10^{-10}$ uncertainty in $a_{\tau}^{\mysmall \rm QED}$.

\subsection{Three-loop Corrections}
\label{subsec:QED3}

More than one hundred diagrams are involved in the evaluation of the
three-loop (sixth-order) {\small QED} contribution. Their analytic
computation required approximately three decades, ending in the late 1990s.
The coefficient $A_1^{(6)}$ arises from 72 diagrams. Its exact expression,
mainly due to Remiddi and his collaborators, reads:\cite{Remiddi,LR96}
\bea 
     A_1^{(6)} &=&  \frac{83}{72} \pi^2 \zeta(3) - \frac{215}{24}
     \zeta(5)  - \frac{239}{2160} \pi^4 + \frac{28259}{5184} +
     \nonumber \\ 
     && + \frac{139}{18} \zeta(3) - \frac{298}{9} \pi^2 \ln2 +
     \frac{17101}{810} \pi^2 +
     \nonumber \\ 
     && + \frac{100}{3} \left[{\rm Li}_4(1/2) + \frac{1}{24} \left( \ln^2
     2 -\pi^2 \right) \ln^2 2 \right]
     \nonumber \\ 
     &=& 1.181 \, 241 \, 456 \, 587 \ldots.
\label{eq:A16}
\eea
This value is in very good agreement with previous
results obtained with numerical methods.\cite{Ki90-95}

The calculation of the exact expression for the coefficient
$A_2^{(6)}(m_l/m_j)$ for arbitrary values of the mass ratio $m_l/m_j$ was
completed in 1993 by Laporta and Remiddi~\cite{La93,LR93} (earlier works
include Refs.~\refcite{A26early}).  Let us focus on $a_{\tau}^{\mysmall \rm
QED}$ ($l=\tau$, $j=e,\mu$). This coefficient can be further split into two
parts: the first one, $A_2^{(6)}(m_l/m_j,\mbox{vac})$, receives
contributions from 36 diagrams containing either electron or muon vacuum
polarization loops,\cite{La93} whereas the second one,
$A_2^{(6)}(m_l/m_j,\mbox{lbl})$, is due to 12 light-by-light scattering
diagrams with either electron or muon loops.\cite{LR93} The exact
expressions for these coefficients are rather complicated, containing
hundreds of polylogarithmic functions up to fifth degree (for the
light-by-light diagrams) and complex arguments (for the vacuum polarization
ones) -- they also involve harmonic polylogarithms.\cite{HarmPol} Series
expansions were provided in Ref.~\refcite{LR93} for the cases of physical
relevance.

Using the recommended mass ratios given in Eqs.~(\ref{eq:rte}) and
(\ref{eq:rtm}), the following values were recently computed from the full
analytic expressions:\cite{MP06}
\bea
     A_2^{(6)}(m_{\tau}/m_{e},\mbox{vac}) &=&  7.256 \, 99 \,(41)
\label{eq:TA26evac}
     \\
     A_2^{(6)}(m_{\tau}/m_{e},\mbox{lbl}) &=&  39.1351 \,(11)
\label{eq:TA26elbl}
     \\
     A_2^{(6)}(m_{\tau}/m_{\mu},\mbox{vac})&=& -0.023 \, 554 \,(51) 
\label{eq:TA26mvac}
     \\
     A_2^{(6)}(m_{\tau}/m_{\mu},\mbox{lbl})&=&  7.033 \, 76 \,(71).
\label{eq:TA26mlbl}
\eea
Almost identical values were obtained employing the approximate series
expansions of Ref.~\refcite{LR93}:
7.25699(41),
39.1351(11),
$-$0.023564(51), 
7.03375(71).\cite{MP06}
The previous estimates of Ref.~\refcite{Narison01} were different: 10.0002,
39.5217, 2.9340, and 4.4412 (no error estimates were provided),
respectively; they are superseded by the results in
Eqs.~(\ref{eq:TA26evac})--(\ref{eq:TA26mlbl}), derived from the exact
analytic expressions. The estimates of Ref.~\refcite{Samuel_tau} compare
slightly better: 7.2670, 39.6, $-0.1222$, 4.47 (no errors provided). In the
specific case of $A_2^{(6)}(m_{\tau}/m_{\mu},\mbox{lbl})$, the values of
Refs.~\refcite{Narison01} and \refcite{Samuel_tau} differ from \eq{TA26mlbl}
because their derivations did not include terms of $O(m_{\mu}/m_{\tau})$,
which turn out to be unexpectedly large.  The sums of
Eqs.~(\ref{eq:TA26evac})--(\ref{eq:TA26elbl}) and
(\ref{eq:TA26mvac})--(\ref{eq:TA26mlbl}) are
\bea
     A_2^{(6)}(m_{\tau}/m_e) &=& 46.3921 \,(15), 
\label{eq:TA26e}
     \\
     A_2^{(6)}(m_{\tau}/m_{\mu})&=& 7.010 \, 21 \,(76).
\label{eq:TA26m}
\eea

The contribution of the three-loop diagrams with both electron- and
muon-loop insertions in the photon propagator was calculated numerically
from the integral expressions of Ref.~\refcite{Samuel91},
obtaining:\cite{MP06}
\be
     A_3^{(6)}(m_{\tau}/m_e,m_{\tau}/m_{\mu}) = 3.347 \, 97 \,(41).
\label{eq:TA36}
\ee
This value disagrees with the results of Refs.~\refcite{Samuel_tau} (1.679)
and \refcite{Narison01} (2.75316). Combining the three-loop results of
Eqs.~(\ref{eq:A16}), (\ref{eq:TA26e}), (\ref{eq:TA26m}) and (\ref{eq:TA36})
one finds the sixth-order {\small QED} coefficient,\cite{MP06}
\be
      C_{\tau}^{(6)}  =  57.9315 \,(27).
\label{eq:TC3}
\ee
The error $2.7 \!\times\! 10^{-3}$ induces a $3 \! \times \!  10^{-11}$
uncertainty in $a_{\tau}^{\mysmall \rm QED}$. The order of magnitude of the
three-loop contribution to $a_{\tau}^{\mysmall \rm QED}$, dominated by the
mass-dependent terms, is comparable to that of {\small EW} and hadronic
effects (see later).

Contrary to the case of the electron and muon $g$$-$$2$, {\small QED}
contributions of order higher than three are not known.\cite{QED_4E,QED45}
(An exception is the mass- and flavor-independent term
$A_4^{(8)}$,\cite{QED_4E} which is however expected to be a very small part
of the complete four-loop contribution.)  Adding up all the above
contributions and using the new value of $\alpha$ derived in
Refs.~\refcite{Gabrielse_a_2006} and \refcite{MP06},
$\alpha^{-1} \, = \,  137.035 \, 999 \, 709 \, (96)$,
one obtains the total {\small QED} contribution to the $g$$-$$2$ of the
$\tau$ lepton,\cite{MP06}
\be
    a_{\tau}^{\mysmall \rm QED} =
    117 \, 324 \, (2) \times 10^{-8}.
\label{eq:TQED}
\ee
The error $\delta a_{\tau}^{\mysmall \rm QED}$ is the uncertainty
$\delta C_{\tau}^{(8)}(\alpha/\pi)^4 \sim \pi^2 \ln^2(m_{\tau}/m_e)
(\alpha/\pi)^4 \sim 2\times 10^{-8}$
assigned to $a_{\tau}^{\mysmall \rm QED}$ for uncalculated four-loop
contributions. As we mentioned earlier, the errors due to the uncertainties
of the $O(\alpha^2)$ and $O(\alpha^3)$ terms are negligible.  The error
induced by the uncertainty of $\alpha$ is only $8 \times 10^{-13}$ (and thus
totally negligible).

\section{Electroweak Contribution to $\at$}
\label{sec:EW}

With respect to Schwinger's contribution, the {\small EW}
correction to the anomalous magnetic moment of the $\tau$ lepton is
suppressed by the ratio $(m_{\tau}/\mw)^2$, where $\mw$ is the mass of the
$W$ boson. Numerically, this contribution is of the same order of magnitude
as the three-loop {\small QED} one.

\subsection{One-loop Contribution}
\label{subsec:EW1}

\begin{figure}[th]
\centerline{\psfig{file=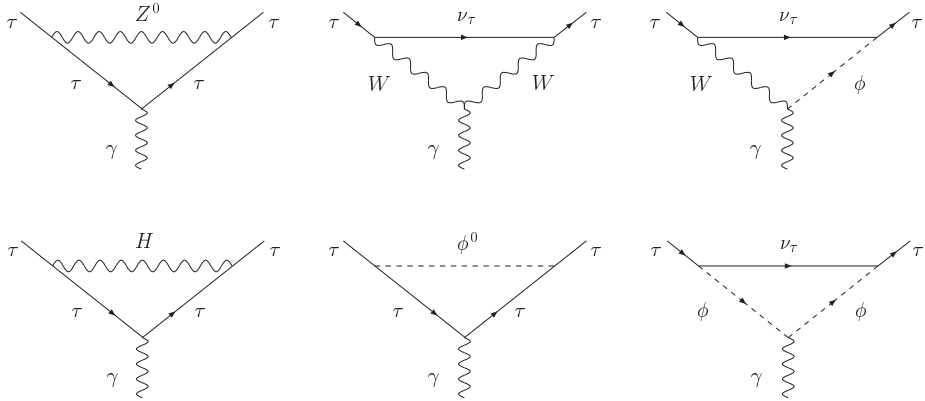,width=5in}}
\vspace*{8pt}
\caption{One-loop electroweak contributions to $a_{\tau}$. The diagram with
  a $W$ and a Goldstone boson ($\phi$) must be counted twice.}
\label{fig:ew1}
\end{figure}

The analytic expression for the one-loop {\small EW} contribution to $\at$,
due to the diagrams in Fig.~\ref{fig:ew1}, is:\cite{ew1loop}
\be
     \at^{\mysmall \rm EW} (\mbox{1 loop}) =
     \frac{5 G_{\mu} m^2_{\tau}}{24 \sqrt{2} \pi^2}
     \left[ 1+ \frac{1}{5}\left(1-4\sin^2\!\theta_{\mysmall{W}}\right)^2 
       + O \left( \frac{m^2_{\tau}}{M^2_{\mysmall{Z,W,H}}} \right) \right],
\label{eq:EWoneloop}
\ee
where $G_{\mu}=1.16637(1) \times 10^{-5}\gev^{-2}$ is the Fermi coupling
constant,\cite{PDG06} $\mz$, $\mw$ and $\mh$ are the masses of the $Z$, $W$
and Higgs bosons, and $\theta_{\mysmall{W}}$ is the weak mixing angle.
Closed analytic expressions for $\at^{\mysmall \rm EW} (\mbox{1 loop})$
taking exactly into account the $m^2_{\tau}/M^2_{\mysmall{B}}$ dependence
($B=Z,W,$ Higgs, or other hypothetical bosons) can be found in
Refs.~\refcite{Studenikin}.  Following Ref.~\refcite{CMV03}, we employ for
$\sin^2\!\theta_{\mysmall{W}}$ the on-shell definition,\cite{Si80}
$\sin^2\!\theta_{\mysmall{W}} = 1-M^2_{\mysmall{W}}/M^2_{\mysmall{Z}}$,
where $\mz=91.1876(21)\gev$,\cite{PDG06} and $\mw$ is the theoretical
{\small SM} prediction of the $W$ mass. The latter can be easily derived
from the simple analytic formulae of Ref.~\refcite{FOPS} (see also
Refs.~\refcite{Formulette}),
\be
      \mw = \left[ 80.3676 
	- 0.05738 \, \ln \! \left(\frac{\mh}{100\gev}\right) -
	0.00892\, \ln^2 \! 
	  \left(\frac{\mh}{100\gev}\right)\right]\!\!\gev,
\label{eq:fops}
\ee
(on-shell scheme {\small II} with $\Delta
\alpha_h^{(5)}=0.02758(35)$,\cite{Burkhardt:2005se}
$\alpha_s(\mz)=0.118(2)$,\cite{PDG06} and $M_{\rm\scriptstyle
top}=171.4(2.1)$ GeV,\cite{newTOP}) leading to $\mw=80.343\gev$ for
$\mh=150\gev$. This result should be compared with the direct experimental
value $\mw=80.403(29)\gev$,\cite{PDG06} which corresponds to a very small
$\mh$. In any case, these shifts in the $\mw$ prediction induced by the
variation of $\mh$ from 114.4 GeV, the current lower bound at 95\%
confidence level,\cite{LEPHIGGS} up to a few hundred GeV, only
change $\at^{\mysmall \rm EW} (\mbox{1 loop})$ by amounts of
$O(10^{-10})$. From \eq{EWoneloop}, including the tiny
$O(m^2_{\tau}/M^2_{\mysmall{Z,W,H}})$ corrections of
Refs.~\refcite{Studenikin}, for $\mh=150\gev$ we get
\be
    \at^{\mysmall \rm EW} (\mbox{1 loop}) = 55.1(1) \times 10^{-8}.
\label{eq:EWoneloopN}
\ee
The uncertainty encompasses the shifts induced by variations of $\mh$ from
114 GeV up to a few hundred GeV, and the tiny uncertainty due to the error
in $m_{\tau}$.

The estimate of the {\small EW} contribution in Ref.~\refcite{Samuel_tau},
$\at^{\mysmall \rm EW} = 55.60(2) \times 10^{-8}$, obtained from the
one-loop formula (without the small corrections of order
$m^2_{\tau}/M^2_{\mysmall{Z,W,H}}$), is similar to our value in
\eq{EWoneloopN}. However, its uncertainty ($2 \times 10^{-10}$) is too
small, and it doesn't contain the two-loop contribution which, as we'll
discuss in the next section, is not negligible.

\subsection{Two-loop Contribution}
\label{subsec:EW2}

The two-loop {\small EW} contributions to $a_l$ ($l\!=\!e$, $\mu$ or $\tau$)
were computed in 1995 by Czarnecki, Krause and Marciano.\cite{CKM95a,CKM95b}
This remarkable calculation leads to a significant reduction of the one-loop
prediction. Na\"{\i}vely one would expect the two-loop {\small EW}
contribution $\at^{\mysmall \rm EW} (\mbox{2 loop})$ to be of order
$(\alpha/\pi) \times \at^{\mysmall \rm EW} (\mbox{1 loop})$, but this turns
out not to be so. As first noticed in the early 1990s,\cite{KKSS}
$a_l^{\mysmall \rm EW} (\mbox{2 loop})$ is actually quite substantial
because of the appearance of terms enhanced by a factor of
$\ln(M_{\mysmall{Z,W}}/m_f)$, where $m_f$ is a fermion mass scale much
smaller than $\mw$.

The two-loop contribution to $\at^{\mysmall \rm EW}$ involves 1678 diagrams
in the linear 't Hooft-Feynman gauge\cite{Kaneko95} (as a check, the authors
of Refs.~\refcite{CKM95a} and \refcite{CKM95b} employed both this gauge and
a nonlinear one in which the vertex of the photon, the $W$ and the
unphysical charged scalar vanishes). It can be divided into fermionic and
bosonic parts; the former, $\at^{\mysmall \rm EW}(\mbox{2 loop fer})$,
includes all two-loop {\small EW} corrections containing closed fermion
loops, whereas all other contributions are grouped into the latter,
$\at^{\mysmall \rm EW}(\mbox{2 loop bos})$. The expressions of
Ref.~\refcite{CKM95b} for the bosonic part were obtained in the
approximation $\mh \! \gg \! M_{\mysmall{W,Z}}$, computing the first two
terms in the expansion in $M^2_{\mysmall{W,Z}}/\mh^2$, and expanding in
$\sin^2\!\theta_{\mysmall{W}}$, keeping the first four terms in this
expansion (this number of powers is sufficient to obtain an exact
coefficient of the large logarithms $\ln(M_{\mysmall{W}}^2/m_{\tau}^2)$).
Recent analyses of the {\small EW} bosonic corrections of the $g$$-$$2$ of
the muon\cite{HSW04-GC05} relaxed these approximations, providing analytic
results valid also for a light Higgs. Considering the present $\mh\!>\!
114.4$ GeV lower bound,\cite{LEPHIGGS} we can safely employ the results of
Ref.~\refcite{CKM95b}, obtaining, for $\mh=150\gev$
$\at^{\mysmall \rm EW}(\mbox{2 loop bos}) = -3.06 \times 10^{-8}$. 
The neglected terms are of $O(10^{-9})$.

\begin{figure}[th]
\centerline{\psfig{file=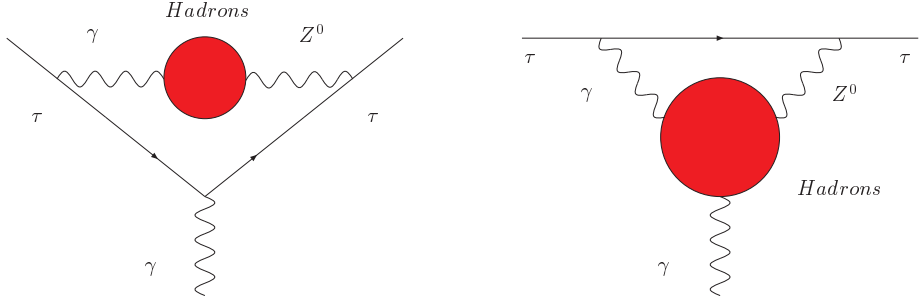,width=5in}}
\vspace*{8pt}
\caption{Some of the fermion-loop diagrams contributing to the $\tau$
  anomalous magnetic moment.}
\label{fig:ew2}
\end{figure}

The fermionic part of $\at^{\mysmall \rm EW}(\mbox{2 loop})$ contains the
contribution of diagrams with light quarks; they involve long-distance
{\small QCD} for which perturbation theory cannot be employed. In
particular, these hadronic uncertainties arise from two types of two-loop
diagrams: the hadronic photon--$Z$ mixing, and quark triangle loops with the
external photon, a virtual photon and a $Z$ attached to them (see
Fig.~\ref{fig:ew2}).
The hadronic uncertainties mainly arise from the latter ones. Two approaches
were suggested for their study: in Ref.~\refcite{CKM95a} the nonperturbative
effects where modeled introducing effective quark masses as a simple way to
account for strong interactions. In view of the high experimental precision
of the $g$$-$$2$ of the muon, a more realistic treatment of the relevant
hadronic dynamics was introduced in Ref.~\refcite{PPD95} within a low-energy
effective field theory approach, later on developed in the detailed analyses
of Refs.~\refcite{CMV03,KPPD02,CMV03b}. However, from a numerical point of
view, the discrepancy between the results provided by these two different
approaches turns out to be irrelevant for the present interpretation of the
experimental result of the muon $g$$-$$2$, in spite of its precision. The
use of effective quark masses for the study of the $g$$-$$2$ of the $\tau$
-- whose experimental precision is a far cry from that of the muon! -- thus
appears to be sufficient at present. The tiny hadronic $\gamma$--$Z$ mixing
terms can be evaluated either in the free quark approximation or via a
dispersion relation using data from $e^+e^-$ annihilation into hadrons; the
difference was shown to be numerically insignificant.\cite{CMV03}

References \refcite{CKM95a} and \refcite{KKSS} contain simple approximate
expressions for the contributions of the diagrams with fermion triangle
loops shown in Fig.~\ref{fig:ew2} (right). In general, for a lepton
$l\!=\!e$, $\mu$ or $\tau$, neglecting small mass-ratios, they are
\be a_{l}^{\mysmall \rm EW}(\mbox{2 loop fer; triangle loops}) \simeq
       \sum_{f} \frac{5 G_{\mu} m^2_l}{24 \sqrt{2} \pi^2}
       \left[\frac{\alpha}{\pi} \frac{18}{5} N_f I_3^f Q_f^2 \Delta C(f)
       \right],
\label{eq:EW2LFd}
\ee
where $I_3^f$ is the third component of the weak isospin of the fermion $f$
in the loop, $Q_f$ is its charge, $N_f$ its number of colors (3 for quarks,
1 for leptons), and
\be 
        \Delta C(f) \simeq \left\{ \begin{array}{ll}
	    \ln (\mz^2/m_l^2) + 5/6 
	    & \textrm{~~~if~~ $m_f \ll m_l \ll \mz$} \\
	    \ln (\mz^2/m_l^2) - 8 \pi^2/27 + 11/18
	    & \textrm{~~~if~~ $m_f = m_l \ll \mz$} \\
	    \ln (\mz^2/m_f^2) -2
	    & \textrm{~~~if~~ $m_l \ll m_f \ll \mz$} \\
	    ~\!\![(\mz^2/M_{\rm\scriptstyle top}^2)/6 - 2/3]
	    \ln (M_{\rm\scriptstyle top}^2/\mz^2) \,\,+ \\
	    +\, (5/18)(\mz^2/M_{\rm\scriptstyle top}^2) -4/3
	    & \textrm{~~~if~~ $m_f = M_{\rm\scriptstyle top}$.}
	    \end{array} \right. 
\label{eq:EW2LFdlogs}
\ee
The contribution of the top-quark triangle loop diagram of
Fig.~\ref{fig:ew2} (right) with the $Z$ boson replaced by the neutral
Goldstone boson ($\phi^0$) has also been included in this
expression.\cite{CKM95a}
It is clear from Eqs.~(\ref{eq:EW2LFd})--(\ref{eq:EW2LFdlogs}) that the
logarithms $\ln(\mz)$ cancel in sums over all fermions of a given
generation, as long as $m_f \!\ll\! \mz$, due to the no-anomaly condition
$\sum_f N_f I_3^f Q_f^2 \!=0\!$ valid within every
generation.\cite{CKM95a,PPD95} This does not occur for the third generation
due to the large mass of the top quark. Note that this short-distance
cancellation does not get modified by strong interaction effects on the
quark triangle diagrams.\cite{CMV03,V2002}

Contrary to the case of the muon $g$$-$$2$, where all fermion masses, with
the exception of $m_e$, enter in $a_{\mu}^{\mysmall \rm EW}(\mbox{2 loop
fer; triangle loops})$, the approximate expressions in \eq{EW2LFdlogs} show
that this is not the case for the $g$$-$$2$ of the $\tau$ lepton. Indeed,
due to the high infrared cut-off set by $m_{\tau}$, \eq{EW2LFd} for
$l\!=\!\tau$ does not depend on any fermion mass lighter than $m_{\tau}$;
apart from $m_{\tau}$, it only depends on $M_{\rm\scriptstyle top}$ and
$m_{b}$, the masses of the top and bottom quarks (assuming
$m_c\!<\!m_{\tau}$). The charm contribution requires some care, as the crude
approximation provided by \eq{EW2LFdlogs} for a charm lighter than the
$\tau$ lepton is valid only if $m_c \!\ll \!m_{\tau}$. Clearly, this is not
a good approximation, and the spurious shift induced by \eq{EW2LFdlogs} when
$m_c$ is varied across the $m_{\tau}$ threshold is of $O(10^{-8})$. One
possibility is to use \eq{EW2LFdlogs} with $m_c$ equal to
$m_{\tau}$.\cite{CKM95b,CK96} Better still, we numerically integrated the
exact expressions for $\Delta C(f)$ provided in Ref.~\refcite{KKS} for
arbitrary values of $m_f$, obtaining a smooth dependence on the value of
$m_c$. For completeness we repeated this detailed analysis for all light
fermions. As expected, the result depends very mildly on the values chosen
for their masses.  Employing the values $m_u\!=\!m_d\!=\!0.3$ GeV,
$m_s\!=\!0.5$ GeV, $m_c\!=\!1.5$ GeV and $m_b\!=\!4.5$ GeV, and adding to
\eq{EW2LFd} the contribution of the remaining fermionic two-loop diagrams
studied in Ref.~\refcite{CKM95a}, for $\mh\!=\!150\gev$ we obtain
$\at^{\mysmall \rm EW}(\mbox{2 loop fer}) = -4.68 \times 10^{-8}$.
In this evaluation we also included the tiny $O(10^{-9})$ contribution of
the $\gamma$--$Z$ mixing diagrams, suppressed by
($1-4\sin^2\!\theta_{\mysmall{W}}) \!\sim\! 0.1$ for quarks and
($1-4\sin^2\!\theta_{\mysmall{W}})^2$ for leptons, via the explicit formulae
of Ref.~\refcite{CMV03}.

The sum of the fermionic and bosonic two-loop {\small EW} contributions
described above gives
$\at^{\mysmall \rm EW}(\mbox{2 loop}) \!=\! -7.74 \times 10^{-8}$, 
a 14\% reduction of the one-loop result.  The leading-logarithm three-loop
{\small EW} contributions to the muon $g$$-$$2$ were determined to be
extremely small via renormalization-group analyses.\cite{CMV03,DGi98} We
assigned to our $\tau$ lepton $g$$-$$2$ {\small EW} result an additional
uncertainty of
$O[\at^{\mysmall \rm EW}(\mbox{2 loop}) (\alpha/\pi)
\ln(\mz^2/m_{\tau}^2)] \!\sim\! O(10^{-9})$
to account for these neglected three-loop effects. Adding $\at^{\mysmall \rm
EW}(\mbox{2 loop})$ to the one-loop value of \eq{EWoneloopN} we get our
total {\small EW} correction (for $\mh\!=\!150\gev$)
\be
    \at^{\mysmall \rm EW} = 47.4 (5) \times 10^{-8}.
\label{eq:TEW}
\ee
The uncertainty allows $\mh$ to range from 114 GeV up to $\sim \! 300$ GeV,
and reflects the estimated errors induced by hadronic loop effects ($m_u$
and $m_d$ can vary between 70 MeV and 400 MeV), neglected two-loop bosonic
terms, and the missing three-loop contribution. It also includes the small
errors due to the uncertainties in $M_{\rm\scriptstyle top}$ and
$m_{\tau}$. The value in \eq{TEW} is in agreement with the prediction
  $\at^{\mysmall \rm EW} = 47 (1) \times 10^{-8}$,\cite{CK96,Narison01}
with a reduced uncertainty.  As we mentioned in Sec.~\ref{subsec:EW1}, the
{\small EW} estimate of Ref.~\refcite{Samuel_tau},
$\at^{\mysmall \rm EW} = 55.60(2) \times 10^{-8}$, mainly differs from
\eq{TEW} in that it doesn't include the two-loop corrections.

\section{The Hadronic Contribution}
\label{sec:HAD}

In this section we will analyze $a_{\tau}^{\mysmall \rm HAD}$, the
contribution to the $\tau$ anomalous magnetic moment arising from {\small
QED} diagrams involving hadrons. Hadronic effects in (two-loop) {\small EW}
contributions are already included in $\at^{\mysmall \rm EW}$ (see the
previous section).

\subsection{Leading-order Hadronic Contribution}
\label{subsec:HLO}

\begin{figure}[th]
\centerline{\psfig{file=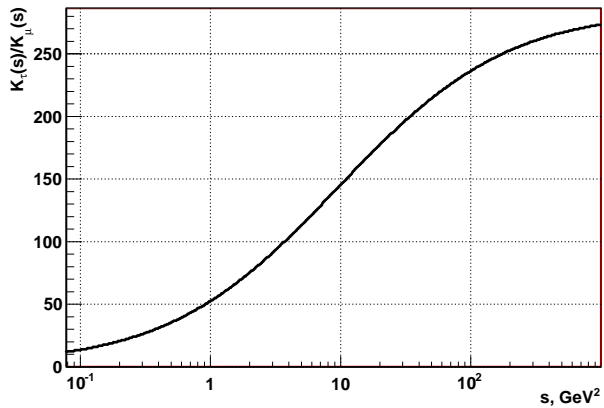,width=0.7\textwidth}}
\vspace*{8pt}
\caption{Ratio of the kernels $K_{\tau}(s)/K_{\mu}(s)$.}
\label{fig:rat}
\end{figure}

Similarly to the case of the muon $g$$-$$2$, the leading-order hadronic
contribution to the $\tau$ lepton anomalous magnetic moment is given by the
dispersion integral:\cite{DispInt}
\begin{equation}
{a}^{\mysmall \rm HLO}_{\tau}=
 \frac{m^2_\tau}{12\pi^3}
\int\limits_{4m^2_\pi}^{\infty} ds\: 
\frac{\sigma^{(0)}(e^+e^- \to {\rm hadrons})\:K_{\tau}(s)}{s},
\label{eq:dispint}
\end{equation}
where the kernel $K_{\tau}(s)$ is a bounded function of energy monotonously
increasing to unity at $s \to \infty$, and $\sigma^{(0)}(e^+e^- \to {\rm
hadrons})$ is the total hadronic cross section of the $e^+e^-$ annihilation
in the Born approximation.  In Fig.~\ref{fig:rat} we plot the ratio of the
kernels in the $\tau$ lepton and muon case.  Clearly, although the role of
the low energies is still very important, the different structure of
$K_{\tau}$ compared to $K_{\mu}$, induced by the higher mass of the $\tau$,
results in a relatively higher role of the larger energies.

The history of these calculations is not as rich as that of the muon. The
first calculation performed in 1978 in Ref.~\refcite{nar78} was based on
experimental data available at that time below 7.4~GeV, whereas at higher
energies the asymptotic {\small QCD} prediction was used.  Ten years
later, a rough estimate was made in Ref.~\refcite{bs88} based on low energy
$e^+e^-$ data.  In Ref.~\refcite{Samuel_tau} the contribution of the $\rho$
meson was estimated by integrating the approximation obtained using the
Breit-Wigner curve, while other contributions used the data. The accuracy of
the calculation was considerably improved in Refs.~\refcite{ej95,j96} where,
below 40~GeV, only data were used. In Ref.~\refcite{Narison01}, data were
only used below 3~GeV (together with the experimental parameters of the
$J/\psi$ and $\Upsilon$ family states). In our opinion this can
significantly underestimate the resulting uncertainty. In addition, in the
same reference, data from $\tau$ lepton decays were extensively used; as it
is known today, this leads to higher spectral functions than in $e^+e^-$
case,\cite{dehz1,dehz2} and can therefore overestimate the result. The
results of these calculations are summarized in Table~\ref{tab:atau}.
%
\begin{table}[h]
\tbl{Calculations of the leading-order hadronic part of ${a}_{\tau}$.}
{\begin{tabular}{@{}lcc@{}} 
\toprule
Author & Year & ${\rm a}_{\tau}^{\mysmall \rm HLO} \times 10^{8}$  \\
\colrule
S.~Narison \cite{nar78} & 1978 & 370 $\pm$ 40 \\
B.C.~Barish and R.~Stroynowski~\cite{bs88} & 1988 & $\sim 350$ \\
M.A.~Samuel et al. \cite{Samuel_tau} & 1991 &  $360 \pm 32$ \\
S.~Eidelman and F.~Jegerlehner \cite{ej95,j96} & 1995 & 
$338.4 \pm 2.0 \pm 9.1$ \\
S.~Narison \cite{Narison01} & 2001 & 353.6 $\pm$ 4.0 \\
This work                   & 2007 & 337.5 $\pm$ 3.7 \\
\hline
M.~Benmerrouche et al.~\cite{ben93} & 1993 & 197--246          \\
F.~Hamzeh and N.F.~Nasrallah \cite{hn93} & 1993 & $280 \pm 20$ \\
B.~Holdom et al.~\cite{hold} & 1994 & $320 \pm 10$ \\
A.E.~Dorokhov~\cite{dor} & 2005 & $310 \pm 20$ \\
\botrule
\end{tabular}}
\label{tab:atau}
\end{table}

For completeness, in the second part of Table~\ref{tab:atau} we also show
purely theoretical estimates. The analysis based on {\small QCD} sum rules
performed in Ref.~\refcite{ben93} gives results which strongly depend on the
choice of quark and gluon condensates. {\small QCD} sum rules are also used
in Ref.~\refcite{hn93}.  In Ref.~\refcite{hold} the authors use a nonlocal
constituent quark model for the description of the photon vacuum
polarization function $\Pi_{\rm had}(q^2)$ at space-like momenta and obtain
$a_{\tau}^{\mysmall \rm HLO} = 3.2(1) \times 10^{-6}$, close to the
estimates based on the experimental data. They also show that a simpler
model with constituent quark masses independent of momentum is strongly
dependent on the values chosen for the quark masses. For example, with
$m_u=m_d=330$~MeV and $m_s=$~550~MeV their result is $a_{\tau}^{\mysmall \rm
HLO} = 2.2(1) \times 10^{-6}$, i.e., significantly smaller than the previous
estimate. They could reproduce the value $a_{\tau}^{\mysmall \rm HLO} =
3.2(1) \times 10^{-6}$ using $m_u=m_d=m_s=201$~MeV.  In a recent analysis
using the instanton liquid model the author obtains $(3.1 \pm 0.2) \times
10^{-6}$.\cite{dor} All these estimates somewhat undervalue the hadronic
contribution and have rather large uncertainties.

\begin{figure}[th]
\centerline{\psfig{file=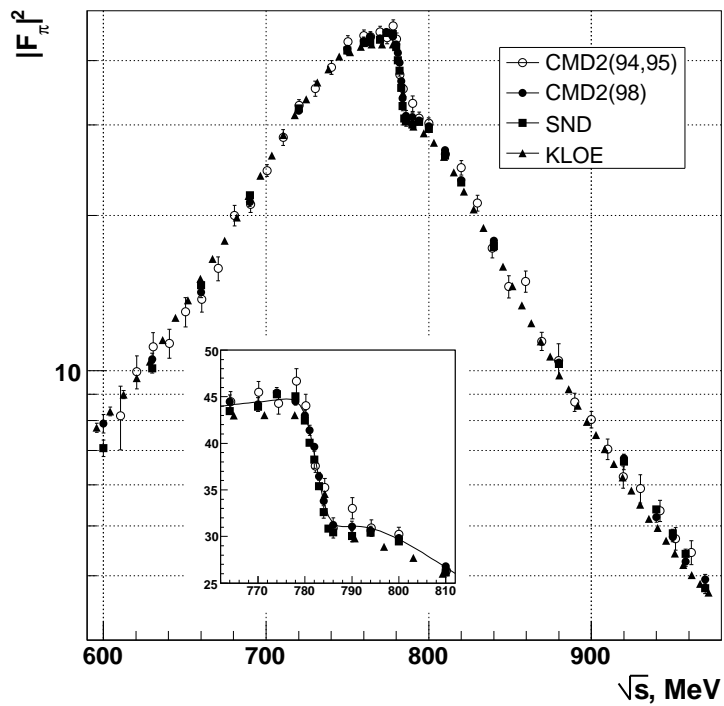,width=0.8\textwidth}}
\vspace*{8pt}
\caption{CMD-2, SND and KLOE data on $|F_{\pi}|^2$ in the $\rho$ meson
  energy range.}
\label{fig:pi}
\end{figure}

We updated the calculation of the leading-order contribution using the whole
bulk of experimental data below 12 GeV, which include old data compiled in
Refs.~\refcite{ej95,dehz1,dehz2}, as well as the recent datasets from the
{\small CMD-2}\cite{cmd2rho,cmd2rest} and {\small
SND}\cite{sndrest,sndff,sndks} experiments in Novosibirsk, and from the
radiative return studies at {\small KLOE} in Frascati\cite{kloe} and BaBar
at {\small SLAC}.\cite{babr} The improvement is particularly visible in the
channel $e^+e^- \to\pi^+\pi^-$, where four new independent measurements
exist in the most important $\rho$ meson region: {\small
CMD-2},\cite{cmd2rho} {\small SND},\cite{sndff} and {\small
KLOE}\cite{kloe} (see Fig.~\ref{fig:pi}). Our result is
\be
    a_{\tau}^{\mysmall \rm HLO} =  337.5 \, (3.7) \times 10^{-8}
\label{eq:THLO}
\ee
(we recently presented a preliminary estimate of this value in
Ref.~\refcite{TAU06}).
The breakdown of the contributions of different energy regions as well
as their relative fractions in the total leading-order contribution are
given in Table~\ref{tab:at}. The contribution of the $\rho$ meson
energy range is still important, but its relative weight is smaller
than in the case of the muon anomaly, 51.3\% compared to about 
72\%.\cite{dehz1,dehz2} The contributions of the narrow resonances
($J/\psi$ and $\Upsilon$ families) are included in the corresponding
energy regions. It is worth noting that uncertainties of the 
contributions from the hadronic continuum are larger than that of the very
precise $2\pi$ one. The overall uncertainty is 2.5 times smaller than
that of the previous data-based prediction.\cite{ej95,j96}    
%
\begin{table}[h]
\tbl{Summary of the contributions to $a_{\tau}^{\mysmall \rm HLO}$ from
  various energy intervals.}
{\begin{tabular}{@{}ccc@{}} 
\toprule
$\sqrt{s}$, GeV & $\Delta{a_{\tau}^{\mysmall \rm HLO}}\!\times 10^{8}$ &
 $\Delta{a_{\tau}^{\mysmall \rm HLO}}$, \% \\
\colrule
 2$\pi, <$ 2 & 173.3 $\pm$ 1.6 & 51.3  \\
 $\omega$ & 15.0 $\pm$ 0.4 & 4.4 \\ 
 $\phi$ & 21.1 $\pm$ 0.4 & 6.3  \\
  0.6--2.0 & 48.2 $\pm$ 2.4 & 14.3 \\
2.0--5.0 & 57.9 $\pm$ 2.0 & 17.2   \\
5.0--12.0 & 16.9 $\pm$ 1.1 & 5.0  \\
 $>$ 12.0 & 5.1 & 1.5 \\
\hline
 Total & 337.5 $\pm$ 3.7 & 100.0 \\
\botrule
\end{tabular}}
\label{tab:at}
\end{table}
%

\subsection{Higher-order Hadronic Contributions}
\label{subsec:HHO}

The hadronic higher-order $(\alpha^3)$ contribution $a_{\tau}^{\mysmall \rm
HHO}$ can be divided into two parts:
$
     a_{\tau}^{\mysmall \rm HHO}=
     a_{\tau}^{\mysmall \rm HHO}(\mbox{vp})+
     a_{\tau}^{\mysmall \rm HHO}(\mbox{lbl}).
$
The first one is the $O(\alpha^3)$ contribution of diagrams containing
hadronic self-energy insertions in the photon propagators.  It was
determined by Krause in 1996:\cite{Krause96} 
\be
a_{\tau}^{\mysmall \rm HHO}(\mbox{vp})= 7.6 (2) \times 10^{-8}.
\label{eq:THHOVAC}
\ee
Note that na\"{\i}vely rescaling the muon result by the factor
$m_{\tau}^2/m_{\mu}^2$ (as it was done in Ref.~\refcite{Samuel_tau}) leads
to the totally incorrect estimate $a_{\tau}^{\mysmall \rm HHO}(\mbox{vp})=
(-101\times 10^{-11}) \times m_{\tau}^2/m_{\mu}^2 = -29 \times 10^{-8}$ (the
$a_{\mu}^{\mysmall \rm HHO}(\mbox{vp})$ value is from
Ref.~\refcite{Krause96}); even the sign is wrong!

The second term, also of $O(\alpha^3)$, is the hadronic light-by-light
contribution. Similarly to the case of the muon $g$$-$$2$, this term cannot
be directly determined via a dispersion relation approach using data (unlike
the leading-order hadronic contribution), and its evaluation therefore
relies on specific models of low-energy hadronic interactions with
electromagnetic currents. Actually, very few estimates of
$a_{\tau}^{\mbox{$\scriptscriptstyle{\rm HHO}$}}(\mbox{lbl})$ exist in the
literature,\cite{Narison01,Samuel_tau,Krause96} and all of them were
obtained simply rescaling the muon results $a_{\mu}^{\mysmall \rm
HHO}(\mbox{lbl})$ by a factor $m_{\tau}^2/m_{\mu}^2$.
Following this very na\"{\i}ve procedure, the $a_{\tau}^{\mysmall \rm
HHO}(\mbox{lbl})$ estimate varies between
$     a_{\tau}^{\mysmall \rm HHO}(\mbox{lbl}) = 
[80(40) \times 10^{-11}]\times (m_{\tau}^2/m_{\mu}^2)= 23(11)\times 10^{-8}
$, 
and
$     a_{\tau}^{\mysmall \rm HHO}(\mbox{lbl}) = 
[136(25) \times 10^{-11}]\times (m_{\tau}^2/m_{\mu}^2)= 38(7)\times 10^{-8}
$, 
according to the values chosen for $a_{\mu}^{\mysmall \rm HHO}(\mbox{lbl})$
from Refs.~\refcite{lbl} and \refcite{MV03}, respectively.

These very na\"{\i}ve estimates fall short of what is needed. Consider the
function $A_2^{(6)}(m_l/m_j,\mbox{lbl})$, the three-loop {\small QED}
contribution to the $g$$-$$2$ of a lepton of mass $m_l$ due to
light-by-light diagrams involving loops of a fermion of mass $m_j$ (see
Sec.~\ref{subsec:QED3}). The exact expression of this function, computed in
Ref.~\refcite{LR93} for arbitrary values of the mass ratio $m_l/m_j$, is
rather complicated, but series expansions were provided in the same article
for the cases of physical relevance. In particular, if $m_j \gg m_l$, then
$A_2^{(6)}(m_l/m_j,\mbox{lbl}) \sim (m_l/m_j)^2$. This implies that, for
example, the (negligible) part of $a_{\tau}^{\mysmall \rm HHO}(\mbox{lbl})$
due to diagrams with a top-quark loop can be reasonably estimated simply
rescaling the corresponding part of $a_{\mu}^{\mysmall \rm HHO}(\mbox{lbl})$
by a factor $m_{\tau}^2/m_{\mu}^2$. On the other hand, to compute the
dominant contributions to $a_{\tau}^{\mysmall \rm HHO}(\mbox{lbl})$, i.e.\
those induced by the light quarks, we need the opposite case: $m_j \ll m_l =
m_{\tau}$. In this limit, $A_2^{(6)}(m_l/m_j,\mbox{lbl})$ does not scale as
$(m_l/m_j)^2$, and a na\"{\i}ve rescaling of $a_{\mu}^{\mysmall \rm
HHO}(\mbox{lbl})$ by $m_{\tau}^2/m_{\mu}^2$ to derive $a_{\tau}^{\mysmall
\rm HHO}(\mbox{lbl})$ leads to an incorrect estimate.

We therefore decided to perform a parton-level estimate of
$a_{\tau}^{\mysmall \rm HHO}(\mbox{lbl})$ based on the exact expression for
$A_2^{(6)}(m_l/m_j,\mbox{lbl})$ using the quark masses recently proposed in
Ref.~\refcite{ES06} for the determination of $a_{\mu}^{\mysmall \rm
HHO}(\mbox{lbl})$: $m_u=m_d=176$ MeV, $m_s=305$ MeV, $m_c=1.18$ GeV and
$m_b=4$ GeV (note that with these values the authors of Ref.~\refcite{ES06}
obtain $a_{\mu}^{\mysmall \rm HHO}(\mbox{lbl}) = 136 \times 10^{-11}$, in
perfect agreement with the value in Ref.~\refcite{MV03} -- see also
Ref.~\refcite{Pivovarov03} for a similar earlier determination). We obtain
\be
a_{\tau}^{\mysmall \rm HHO}(\mbox{lbl})= 5 (3) \times 10^{-8}.
\label{eq:THHOLBL}
\ee
This value is much lower than those obtained by simple rescaling of
$a_{\mu}^{\mysmall \rm HHO}(\mbox{lbl})$ by $m_{\tau}^2/m_{\mu}^2$.  The
up-quark provides the dominating contribution; the uncertainty $\delta
a_{\tau}^{\mysmall \rm HHO}(\mbox{lbl})= 3 \times 10^{-8}$ allows $m_u$ to
range from 70 MeV up to 400 MeV. Further independent studies (following the
approach of Ref.~\refcite{MV03}, for example) would provide an important
check of this result.

The total hadronic contribution to the anomalous magnetic moment of the
$\tau$ lepton can be immediately derived adding the values in
Eqs.~(\ref{eq:THLO}), (\ref{eq:THHOVAC}) and (\ref{eq:THHOLBL}),
\be
      a_{\tau}^{\mysmall \rm HAD} = a_{\tau}^{\mysmall \rm HLO}+
      a_{\tau}^{\mysmall \rm HHO}(\mbox{vp})+
      a_{\tau}^{\mysmall \rm HHO}(\mbox{lbl}) =
      350.1 \, (4.8) \times 10^{-8}.
\label{eq:THAD}
\ee
Errors were added in quadrature.

\section{The Standard Model prediction for \boldmath $a_{\tau}$ \unboldmath}
\label{sec:SM}

We can now add up all the contributions discussed in the previous sections
to derive the {\small SM} prediction for $a_{\tau}$:
\be
    a_{\tau}^{\mysmall \rm SM} = 
         a_{\tau}^{\mysmall \rm QED} +
         a_{\tau}^{\mysmall \rm EW}  +
	 a_{\tau}^{\mysmall \rm HLO}  +
	 a_{\tau}^{\mysmall \rm HHO}(\mbox{vp})  +
	 a_{\tau}^{\mysmall \rm HHO}(\mbox{lbl}),
\label{eq:sm}
\ee
where
$$
\begin{array}{lclr}
      a_{\tau}^{\mysmall \rm QED}   &= & 
       117 \, 324 \, (2)     &\times 10^{-8}        \\  
      a_{\tau}^{\mysmall \rm EW}    &= &
      47.4 \, (5)            &\times 10^{-8}        \\
      a_{\tau}^{\mysmall \rm HLO} &= & 
      337.5 \, (3.7)         &\times 10^{-8}        \\
      a_{\tau}^{\mysmall \rm HHO}(\mbox{vp})\,\,&= & 
      7.6 \, (2)             &\times 10^{-8}        \\
      a_{\tau}^{\mysmall \rm HHO}(\mbox{lbl})\,\,&= &  
      5 \,   (3)             & \times 10^{-8}           
\end{array}
$$
(the sum of the hadronic contributions is given in \eq{THAD}).  Adding
errors in quadrature, our final result is
\be
    a_{\tau}^{\mysmall \rm SM} =  117 \, 721 \, (5) \times 10^{-8}.  
\label{eq:nsm}
\ee
%


The present {\small PDG} limit on the anomalous magnetic moment of the
$\tau$ lepton was derived in 2004 by the {\small DELPHI} collaboration from
$e^+e^- \to e^+e^-\tau^+\tau^-$ total cross section measurements at $\sqrt
s$ between 183 and 208 GeV at {\small LEP2}:\cite{delphi}
\be
                         -0.052 < a_{\tau} < 0.013
\label{eq:exp_delphi1}
\ee
at 95\% confidence level. The authors of Ref.~\refcite{delphi} also quote
their result in the form of central value and error:
\be
                         a_{\tau} = -0.018 (17).
\label{eq:exp_delphi2}
\ee
Comparing this result with \eq{nsm} (their difference is roughly one
standard deviation), it is clear that the sensitivity of the best existing
measurements is still more than an order of magnitude worse than needed. A
reanalysis of various measurements of the cross section of the process
$e^+e^- \to \tau^+\tau^-$, the transverse $\tau$ polarization and asymmetry
at {\small LEP} and {\small SLD}, as well as of the decay width $\Gamma(W
\to \tau\nu_{\tau})$ at {\small LEP} and Tevatron, allowed to set a stronger
model-independent limit:\cite{arcadi}
\be
                       -0.007 < a_{\tau} < 0.005.
\ee
Other limits on $a_{\tau}$ can be found in Refs.~\refcite{at_exp}.

\section{Conclusions}
\label{sec:CONC}

In this article we reviewed and updated the {\small SM} prediction of the
$\tau$ lepton $g$$-$$2$. Updated {\small QED} and electroweak contributions
were presented, together with new values of the leading-order hadronic term,
based on the recent low energy $e^+ e^-$ data from BaBar, {\small CMD-2},
{\small KLOE} and {\small SND}, and of the 
hadronic light-by-light contribution.
These results were confronted in Sec.~\ref{sec:SM} to the available
experimental bounds on the $\tau$ lepton anomaly.

As we already mentioned in the Introduction, quite generally, {\small NP}
associated with a scale $\Lambda$ is expected to modify the {\small SM}
prediction of the anomalous magnetic moment of a lepton $l$ of mass $m_l$ by
a contribution $a_l^{\mysmall \rm NP} \sim m_l^2/\Lambda^2$. Therefore,
given the large factor $m_{\tau}^2/m_{\mu}^2 \sim 283$, the $g$$-$$2$ of the
$\tau$ lepton is much more sensitive than the muon one to {\small EW} and
{\small NP} loop effects that give contributions $\sim m_l^2$, making its
measurement an excellent opportunity to unveil (or just constrain) {\small
NP} effects.

Another interesting feature can be observed comparing the magnitude of the
{\small EW} and hadronic contributions to the muon and $\tau$ lepton
$g$$-$$2$.  The {\small EW} contribution to the $g$$-$$2$ of the $\tau$ is
only a factor of seven smaller than the hadronic one, compared to a factor
of 45 for the $g$$-$$2$ of the muon. Also, while the {\small EW}
contribution to $a_{\mu}^{\mysmall \rm SM}$ is only a factor of three larger
than the present uncertainty of the hadronic contribution, this factor
raises to 10 for the $\tau$ lepton. If a {\small NP} contribution were of
the same order of magnitude as the {\small EW} one, from a purely
theoretical point of view, the $g$$-$$2$ of the $\tau$ would provide a much
cleaner test of the presence (or absence) of such {\small NP} effects than
the muon one. Indeed, if this were the case, such a {\small NP} contribution
to the $\tau$ lepton $g$$-$$2$ would be much larger than the hadronic
uncertainty, which is currently the limiting factor of the {\small SM}
prediction.

Unfortunately, the very short lifetime of the $\tau$ lepton makes it very
difficult to determine its anomalous magnetic moment by measuring its spin
precession in the magnetic field, like in the muon $g$$-$$2$
experiment.\cite{bnl} Instead, experiments focused on high-precision
measurements of the $\tau$ lepton pair production in various high-energy
processes, comparing the measured cross sections with the {\small QED}
predictions.\cite{delphi,at_exp} As we can see from \eq{exp_delphi1}, the
sensitivity of the best existing measurements is still more than an order of
magnitude worse than that required to determine $a_{\tau}$.

Nonetheless, the possibility to improve such a measurement is certainly not
excluded.  For example, it was suggested to determine the $\tau$ lepton $g$
factor taking advantage of the radiation amplitude zero which occurs at the
high-energy end of the lepton distribution in radiative $\tau$
decays.\cite{rad_zero} This method requires a very good energy resolution
and could perhaps be employed at a $\tau$-charm or $B$ factory also
benefiting from the possibility to collect very high statistics.  It is not
clear whether the huge data samples at $B$ factories will result in a
corresponding gain for the limits on $a_{\tau}$. Indeed, {\small LEP}
measurements were rather limited by systematic uncertainties, which were of
the order of 2-3\% for the discussed processes and, until now, experiments
at $B$ factories have not yet reached such a level of accuracy in the
absolute measurements of the total cross sections. However, a search for the
$\tau$ lepton electric dipole moment at Belle\cite{inami} showed that with
the appropriate choice of observables, using full information about events,
the improvement in sensitivity can be proportional to the square root of
luminosity, i.e., determined mainly by statistics. One can hope that this is
also the case with the determination of $a_{\tau}$. A similar method to
study $a_{\tau}$ using radiative $W$ decays and potentially very high data
samples at {\small LHC} was suggested in Ref.~\refcite{samu}.  Yet another
method would use the channeling in a bent crystal similarly to the
suggestion for the measurement of magnetic moments of short-living
baryons.\cite{bent_crystal} This method has been successfully tested by the
E761 collaboration at Fermilab, which measured the magnetic moment of the
$\Sigma^+$ hyperon.\cite{bent_crystal2} In the case of the $\tau$ lepton, it
was suggested to use the decay $B^+ \to \tau^+ \nu_{\tau}$, which would
produce polarized $\tau$ leptons.\cite{Samuel_tau} In 1991, when this
suggestion was published, the idea seemed completely unlikely. However, in
the era of $B$ factories, when the decay $B^+ \to \tau^+ \nu_{\tau}$ is
already observed by the Belle collaboration,\cite{btau} and the possibility
of a Super-$B$ factory is actively discussed, this is no longer a
dream. Even more promising could be the realization of this idea in a
dedicated experiment at a hadron collider with its huge number of $B$ mesons
produced and a more suitable geometry.  We believe that a detailed
feasibility study of such an experiment, as well as further attempts to
improve the accuracy of the theoretical prediction for $a_{\tau}$, are quite
timely.

\section*{Acknowledgments}

We would like to thank M.\ Giacomini and F.V.\ Ignatov for many valuable
comments and collaborations on topics presented in this manuscript.  We are
greatly indebted to A.\ Vainshtein for communications concerning the
hadronic light-by-light contribution to $a_{\tau}$ and to F.\ Jegerlehner
for many fruitful discussions. We are also grateful to K.~Inami for an
interesting discussion on the feasibility of $a_{\tau}$ measurements at $B$
factories.
S.E.\ thanks the Dipartimento di Fisica, Universit\`{a} di Padova and INFN,
Sezione di Padova, where part of this work was done, for its
hospitality. The work of S.E.\ was supported in part by the grants of RFBR
06-02-04018 and 06-02-16156 as well as by the grant of DFG GZ: 436 RUS
113/769/0-2.
M.P.\ is grateful to the Instituto de F\'{\i}sica da Universidade Federal
da Bahia, Brasil, for the hospitality during a visit when this manuscript
was finalized.  The work of M.P.\ was supported in part by the European
Community's Marie Curie Research Training Networks under contracts
MRTN-CT-2004-503369 and
MRTN-CT-2006-035505. All diagrams were drawn with {\tt Jaxodraw}.\cite{JAX}


\end{document}